\documentclass[reprint, amsmath, amssymb, aps]{revtex4-2}

\usepackage{graphicx}% Include figure files
\usepackage{dcolumn}% Align table columns on decimal point
\usepackage{bm}% bold math
\usepackage{xcolor}
\usepackage{hyperref}% add hypertext capabilities
\usepackage{CJKutf8}
\begin{document}
\begin{CJK}{UTF8}{gbsn} 
\preprint{APS/123-QED}

\title{Skin-topological effect in two-dimensional nonreciprocal topological superconductor}

\author{Hong Wang}
\affiliation{School of Physics, MOE Key Laboratory for Non-equilibrium Synthesis and Modulation of Condensed Matter, Xi’an Jiaotong University, Xi’an 710049, China}
\author{Ming Lu}
\thanks{Corresponding author: luming@baqis.ac.cn}
\affiliation{Beijing Academy of Quantum Information Sciences, Beijing 100193, China}
\author{Jie Liu}
\thanks{Corresponding author: jieliuphy@xjtu.edu.cn}
\affiliation{School of Physics, MOE Key Laboratory for Non-equilibrium Synthesis and Modulation of Condensed Matter, Xi’an Jiaotong University, Xi’an 710049, China}
\affiliation{Hefei National Laboratory, Hefei 230088, China}
\date{\today}

\begin{abstract}
We investigate the interplay between non-Hermitian skin effect (NHSE) and topological properties in two-dimensional topological superconductor. Two kinds of non-Hermiticity are considered. The first is the spin-independent non-reciprocal hopping, which respects particle-hole symmetry (PHS) and dictates the bulk as well as topological edge modes localize at opposite corners, manifested as the $Z_2$ NHSE protected by PHS. The direction of the localization can be conveniently characterized by the sign of the winding numbers. The other is the spin-dependent non-reciprocal hopping, where the NHSE is protected by time-reversal symmetry (TRS). The Kramers doublets are localized at opposite corners, namely the TRS protected $Z_2$ NHSE. When apply an external magnetic field, its internal symmetry changes from the symplectic class into the orthogonal class, eliminating the NHSE for the states in the spectrum continuum. For the zero energy states which are isolated, the NHSE still has its effects despite the orthogonality. For the spin-independent case, edge states can be effectively tuned by the direction of Zeeman field, where at certain directions the zero-energy edge state can be free of NHSE and uniformly distributed. Our work paves the way for the study of the interplay between topology and non-Hermiticity in superconducting systems.
\end{abstract}

\maketitle

\section{INTRODUCTION}\label{Sec:I}

Topological superconductors have been extensively studied due to their topologically protected majorana zero modes(MZMs) and have made remarkable progress both theoretically and experimentally \cite{qi2011, Alicea2012, Leijnse2012, Beenakker2013, Sato2017, peng2018, dong2018}. An important feature in topological superconductors is the “bulk-boundary correspondence(BBC)”, that is, the generation of MZMs is protected by the bulk topology. With the further study of topological superconductors, the higher-order topological superconductors have emerged \cite{langbehn2017, liu2018, khalaf2018, lado2018, wang2018, yan2018, zhu2018, wang2018oct, volpez2019, zhu2019, yan2019,franca2019, de2020, wu2020, kheirkhah2020, luo2021, li2021}. Its main characteristics are reflected in a new BBC: a $D$-dimensional $n$th-order topological superconductor with a $(d-n)$ -dimensional gapless boundary state$(n\geq 2)$. In higher-order topological superconductors, topological properties are not only reflected in the bulk, but also at the boundaries \cite{li2021}. For example, for a two-dimensional second-order topological superconductor, when the topological properties of adjacent edges are not equivalent, MZMs appear at the intersection of these two edges, i.e., majorana corner modes(MCMs) \cite{yan2018, zhu2018, wang2018oct}.

Recently, the topological properties of non-Hermitian systems have attracted much attention in condensed matter physics. The non-Hermitian system has non-Hermitian skin effect (NHSE) without Hermitian counterpart, and the non-Hermitian term breaks the traditional BBC, causing the energy spectrum under periodic boundary conditions (PBC) very different from that under open boundary conditions (OBC)\cite{lee2016, shen2018, xiong2018, yao2018Aug, xiao2024, yang2020,song2019Dec, yokomizo2019, deng2019,zhang2020, okuma2020}. In two-dimensional or higher dimensional systems, the interaction between non-Hermitian and topological properties can give rise to many new phenomena. As an important example, hybrid skin-topological states become possible when topological edge states and NHSE are localized in different directions. The hybrid skin-topology effects have been implemented in non-reciprocal hopping systems \cite{Lee2019, kawabata2020, Okugawa2020,fu2021} and gain-loss systems \cite{Li2022}. In addition, there are higher-order NHSEs, which are characterized by NHSE realized in topological edge modes rather than the bulk \cite{kawabata2020, Okugawa2020, fu2021, Zhang2021}. As mentioned before, the higher order topological superconductors have been relatively well studied mainly in the Hermitian cases due to its own physical significance and the potential applications in quantum computation. However, its interactions with the ubiquitous present non-Hermiticity have been rarely studied up to now.

In this paper, we analyze the topological properties and NHSE in two-dimensional $p\pm ip$ topological superconductor systems with two different kinds of nonreciprocal hoppings. The first is the spin-independent nonreciprocal hopping, meaning the nonreciprocal hopping of spin-up and spin-down are the same; the second type is the spin-dependent nonreciprocal hopping, in which the nonreciprocal hopping of spin-up and spin-down occur in opposite ways. When spin-independent nonreciprocal hopping is introduced, the NHSE protected by PHS localizes the bulk eigenstates at two opposite corners of the system. The localization direction of NHSE can be predicted by the sign of winding number. The particles and holes have winding numbers with opposite signs, corresponds to $Z_2$ NHSE protected by PHS\cite{ji2024}. At the same time, the topological edge states will also be affected by the $Z_2$ skin effect, localizing the edge states at opposite corners, namely the “$Z_2$ skin-topological effect”. When the Zeeman field is added, edge states will appear as the angle of the Zeeman field changes. In this case, the edge states with zero energy are uniformly distributed on the edge, while the edge states with non-zero energy are localized in the corner by $Z_2$ NHSE. When spin-dependent non-reciprocal hopping is introduced, the NHSE protected by the TRS also localizes the topological edge states at opposite corners. When the $Z_2$ skin-topological mode appears at one corner, its Kramers partner is localized at the opposite corner. However, this NHSE is strongly suppressed by adding the Zeeman field term, the system changes from the symplectic class into the orthogonal class. As the Zeeman field increases, a topological phase transition occurs, leading to the disappearance of NHSE.

This work is organized as follows. In Sec. \ref{Sec:II}, we introduce the Hamiltonians of two nonreciprocal hopping models. In Sec.\ref{Sec:III}, we study the behavior of skin effect and edge states for different models. In Sec.\ref{Sec:IIIA}, we study spin-independent nonreciprocal hopping model, both bulk states and edge states are affected by $Z_2$ skin effect, which protects PHS. In Sec. \ref{Sec:IIIB}, we focus on the spin-dependent nonreciprocal hopping model, and the edge states are influenced by the $Z_2$ skin effect which is protected by the TRS. If we add Zeeman field, the skin effect is strongly suppressed. We give a brief summary in Sec. \ref{Sec.IV}. 

\begin{figure}[t]
     \includegraphics[scale=0.53]{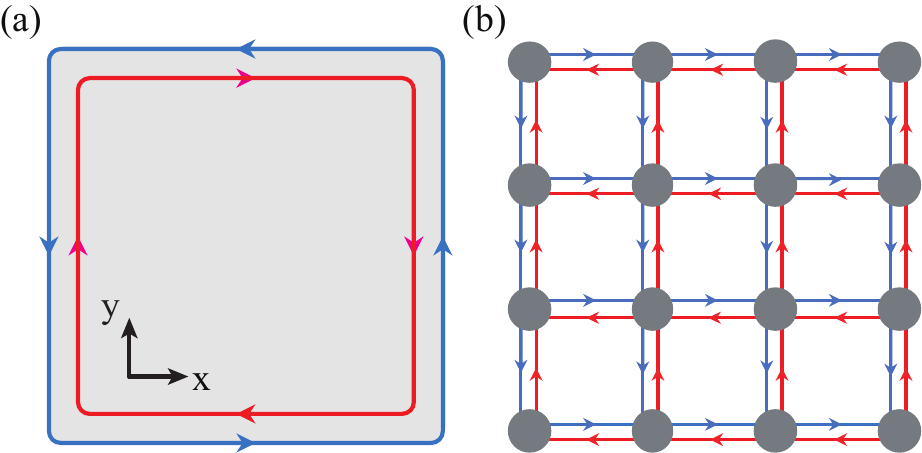}
     \caption{(a)Schematic plot of 2D $p\pm ip$ topological superconductor，which features helical edge state.(b)Tight-binding representation of the model on a square lattice. The red lines$(-t-\delta_i)$ and blue lines$(-t+\delta_i)$  represent nonreciprocal hoppings.}\label{fig1}
\end{figure}

\section{MODEL}\label{Sec:II}

\begin{figure}[t]
     \includegraphics[scale=0.5]{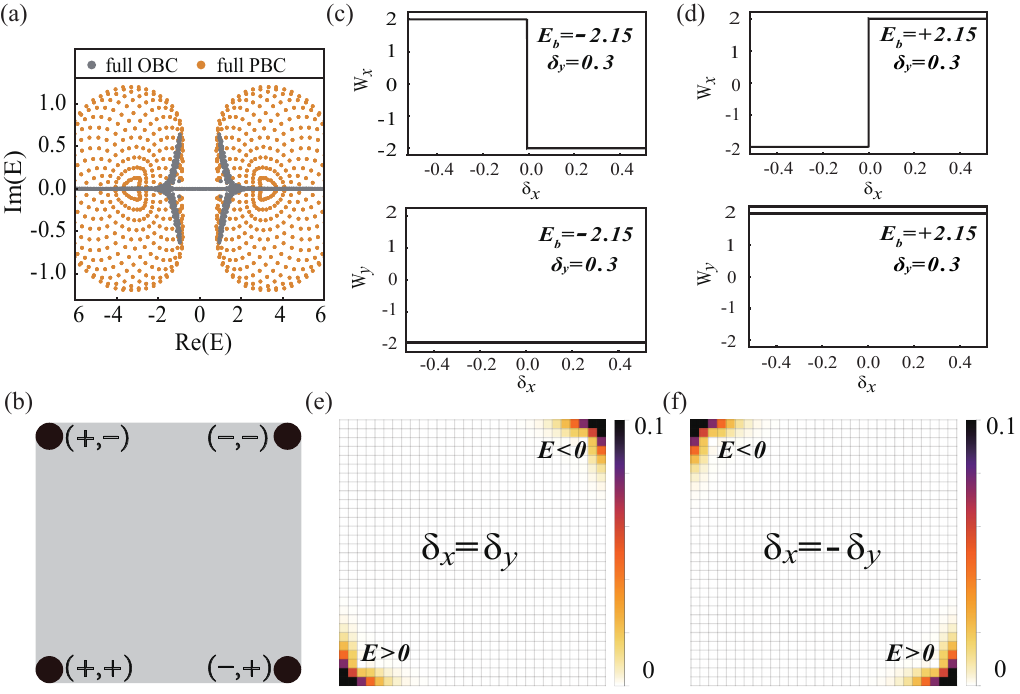}
     \caption{The correspondence between winding number $(W_x,W_y)$ and the position of the skin modes in bulk without Zeeman field. (a)The energy spectrum for non-Hermitian Hamiltonian $H_1(\mathbf{k})$ under the full OBC(gray dots) and full PBC (orange dots), with $\delta_x=0.3$. (b)Correspondence between winding numbers $(W_x,W_y)$ and skin modes. (c)-(d)Winding number with varied $\delta_x$ and fixed $\delta_y=0.3$, in which $k_{y/x}=0$. (e)-(f) Eigenstate distribution of the bulk modes with $\delta_x=\delta_y$ in (e) and $\delta_x=-\delta_y$ in (f).The rest parameters are determined as $\mu=3, t=1, \Delta=1,\delta_y=0.3, V=0$. }\label{fig2}                                                                                                                                                                                                                                                                                                                                                                                                                                                                                                                                                                                                                                                                                                                                                                                                                                                                                                                                                                                                                         
\end{figure}

 We investigate a two-dimensional $p\pm ip$ topological superconductor with different nonreciprocal hoppings. In the Hermitian case, due to the protection of the TRS, a pair of one-dimensional helical edge states appear with opposite spins and opposite directions of motion as shown in Fig. \ref{fig1}(a). When the Zeeman field is added, the edge Hamiltonian will acquire a mass term. When the mass terms of adjacent edges are opposite, MCM will emerge\cite{zhu2018}. The Bloch Hamiltonian is written as
\begin{align}
	H(\mathbf{k}) =\varepsilon(\mathbf{k})\tau_z &-\Delta\tau_x(\sin k_x \sigma_x-\sin k_y \sigma_y)+\mathbf{V}\cdot\bm{\sigma}
\end{align}
where $\tau_i,\sigma_i$ are Pauli matrices in particle-hole and spin spaces, $\varepsilon(\mathbf{k})=\mu-2t(\cos k_x+\cos k_y)$ and $\mathbf{V}=(V \cos\theta,V\sin\theta)$ is the effective Zeeman field with $\theta=\arg(V_x+iV_y)$.

There are two ways to introduce the nonreciprocal hoppings as shown in the Fig. \ref{fig1}(b). In the first case, we add the spin-independent ﻿nonreciprocal hopping, so that the Hamiltonian becomes
\begin{align}
	H_{1}(\mathbf{k}) &=\varepsilon(\mathbf{k})\tau_z-\Delta\tau_x(\sin k_x \sigma_x-\sin k_y \sigma_y)+\mathbf{V}\cdot\mathbf{\sigma}\nonumber\\
	&-2i\delta_x\sin k_x-2i\delta_y\sin k_y	\label{ham:eq1}
\end{align}
where $\delta_i (i=x, y)$ are nonreciprocal hoppings in the $i$ direction. This system respects PHS, $C^{-1}H_1(\mathbf{k})C=-H_{1}^{T}(\mathbf{-k})$ with  $C=\tau_y\sigma_y$.

In the second case, we add the spin-dependent ﻿nonreciprocal hopping, so that the Hamiltonian becomes
\begin{align}
	H_{2}(\mathbf{k}) &=\varepsilon(\mathbf{k})\tau_z-\Delta\tau_x(\sin k_x \sigma_x-\sin k_y \sigma_y)+\mathbf{V}\cdot\mathbf{\sigma}\nonumber\\
	&-2i\delta_x\sin k_x\tau_z\sigma_z-2i\delta_y\sin k_y\tau_{z}\sigma_{z}  \label{ham:eq2}
\end{align}
In the absence of Zeeman fields, the system preserves PHS with $C=\tau_y\sigma_y$ and TRS, $T^{-1}H_2(\mathbf{k})T=H_{2}^{T}(\mathbf{-k})$ with $T=i \sigma_y$. Due to $T T^*=-1$,the Hamiltonian belong to symplectic class. When the Zeeman field is introduced, we have $T'=\tau_z\sigma_x$ satisfying $T'^{-1}H_2(\mathbf{k})T'=H_{2}^{T}(\mathbf{-k})$ and $T'T'^{*}=1$, the Hamiltonian then belong to orthogonal class\cite{Kawabata2020May}.

\begin{figure}[t]
     \includegraphics[scale=0.65]{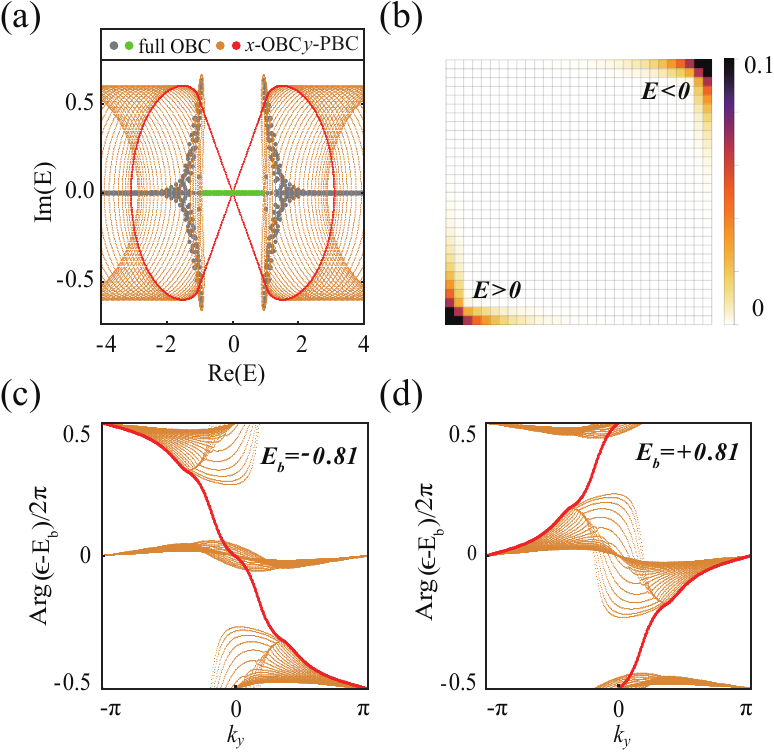}
     \caption{The correspondence between winding number $(\mathcal{W}_x,\mathcal{W}_y)$ and the position of the skin-topological modes  without Zeeman field. (a)The energy spectrum for non-Hermitian Hamiltonian $H_1(\mathbf{k})$ under the full OBC for the bulk (gray) and surface (green) states, the only OBC in the $x$ direction (orange and orange dots), with $V=0, \delta_x=0.3$, the green dots are edge modes. (b) Eigenstate distribution of the edge modes. (c)-(d)Phase plots of the eigenvalues with respect to the wave vector $k_y$. The reference value is $E_b=-0.81$ in (c) and $E_b=0.81$ in (d).}\label{fig3}
\end{figure}

\section{ EDGE STATES AND SKIN EFFECT}\label{Sec:III}

In this section, we discuss the NHSE protected by different symmetries under different nonreciprocal hoppings and their influence on the edge states.

\subsection{\textnormal{\textbf{spin-independent nonreciprocal hopping}}}\label{Sec:IIIA}

Adding spin-independent nonreciprocal hopping as presented in Eq.(\ref{ham:eq1}), the bulk eigenstates and topological edge modes of 2D $p\pm ip$ topological superconductors exhibit NHSE and are located in the corners of the system. We show the energy spectrum for a $30 \times 30$ square lattice in Fig. \ref{fig2}(a) without Zeeman field. The energy spectrum under OBC and PBC conditions is very different in Fig. \ref{fig2}(a), indicating the existence of NHSE. As we know, in one-dimensional system, NHSE under OBC is correlated with the point gap in the PBC energy spectrum, which can be represented by a winding number. For two-dimensional non-Hermitian topological superconductors, NHSE can also be revealed by the point gap topology in cylindrical geometry. The point gap topology can be characterized by the winding number $W(k_\perp,E_b)$ for the reference energy $E_b$, which is defined as
\begin{equation}
	W_\alpha(k_\perp,E_b)=\int^{2\pi}_{0}\frac{dk_\alpha}{2\pi i}\frac{\partial}{\partial k_\alpha}\log \det[H(k_\alpha,k_\perp)-E_b]\label{W:eq3}
\end{equation}
where $\alpha$ indicates $x$ or $y$, $\perp$ is the direction perpendicular to $\alpha$, and $E_b$ is reference energy. If $W_ \alpha(k_\perp,E_b)\neq 0$, then NHSE will occur in the $\alpha$ direction under open boundary conditions. The sign of winding number represents the localization direction of NHSE, and the mapping is shown in the Fig. \ref{fig2}(b). 

Since the system has PHS, the eigenstates of the particles and holes are located on the opposite corner of the system, namely the $Z_2$ skin effect, and the winding number of the particles and holes are opposite to each other:
\begin{align}
	W_\alpha &(k_\perp,E_b)=\int^{2\pi}_{0}\frac{dk_\alpha}{2\pi i}\partial k_\alpha\log \det[H(k_\alpha,k_\perp)-E_b]\nonumber\\
	&=\int^{2\pi}_{0}\frac{dk_\alpha}{2\pi i}\partial k_\alpha\log \det[-\mathcal{C}^{-1}H^{T}(-k_\alpha,-k_\perp)\mathcal{C}-E_b]\nonumber\\ 
	&=\int^{2\pi}_{0}\frac{dk_\alpha}{2\pi i}\partial k_\alpha\log \det[H(-k_\alpha,-k_\perp)+E_b]\nonumber\\
	&=-\int^{2\pi}_{0}\frac{dk_\alpha}{2\pi i}\partial k_\alpha\log \det[H(k_\alpha,-k_\perp)+E_b]\nonumber\\
	&=-W_\alpha(-k_\perp,-E_b).
\end{align}
When $(W_x,W_y)=(-,-)$, the bulk eigenstate resides at the upper right corner, when $(W_x,W_y)=(+,+)$, the bulk eigenstate is localized at the lower left corner, where $+/-$ represent the sign of winding number. For example, for bulk eigenstates with negative energy,  under OBC along $x$ ($x$-OBC) and PBC in the $y$ direction ($y$-PBC)，the winding number $W_x(0,E_b)$ of the Hamiltonian in Eq. (\ref{ham:eq1}) is illustrated in the Fig. \ref{fig2}(c). When $\delta_x=\delta_y$, the negative $W_x$ indicates that the bulk eigenstates with negative energies are localized on the right boundary of the cylinder geometry. Under the $x$-PBC and $y$-OBC, the sign of winding number $W_y(0,E_b)$  is shown to be negative, indicating that the bulk eigenstates with negative energies are localized on the upper boundary of the cylinder geometry. So when $\delta_x=\delta_y$, for the eigenstates of the holes, the winding numbers are $(W_x,W_y)=(-,-)$, these eigenstates are located in the upper right corner; for the eigenstates of the particles, $(W_x,W_y)=(+,+)$ as shown in Fig. \ref{fig2}(d), these are in the lower left corner, as depicted in Fig. \ref{fig2}(e). For $\delta_x=-\delta_y$, the eigenstates of the particles and holes are located in the lower right and upper left corners, respectively, as show in Fig. \ref{fig2}(f).

Apart from the bulk eigenstates affected by $Z_2$ NHSE, the topological edge states are likewise affected, which is denominated as $Z_2$ skin-topological effect. The $Z_2$ skin-topological effect is the interplay between the topological edge states and the $Z_2$ skin effect, which can also be characterized by the point gap topology:
\begin{align}
	\mathcal{W}_\alpha &=\int^{2\pi}_{0}\frac{dk_\alpha}{2\pi i}\partial k_\alpha\log\det[H(k_\alpha)-E_b]\nonumber\\
	&=\sum_{n=1}^{N}\int^{2\pi}_{0}\frac{dk_\alpha}{2\pi i}\partial k_\alpha\arg[E_n(k_\alpha)-E_b]\label{W:eq5}
\end{align}
where $H(k_\alpha)$ is the Hamiltonian in cylindrical geometry, N is the total number of bands,  encompassing both bulk and edge states. As depicted in Fig. \ref{fig3}(a), the energies of the topological edge states under $x$-OBC and $y$-PBC (red dots) circumscribe the energy of the skin-topological states under full OBC(green dots). Due to the presence of PHS, the winding number $\mathcal{W}_y$ of particles and holes are opposite, as illustrated  by Fig. \ref{fig3}(c)-(d), which indicates that the skin-topological modes of particles and holes are located diagonally on the system as shown in Fig. \ref{fig3}(b). For $x$-PBC and $y$-OBC, the winding number $\mathcal{W}_x$ of particles and holes are also reversed. So if the winding numbers are $(\mathcal{W}_x,\mathcal{W}_y)=(-,-)$, the skin-topological modes are located at the lower left corner; if the winding numbers are $(\mathcal{W}_x,\mathcal{W}_y)=(+,+)$, the skin-topological modes are located at the upper right corner.

\begin{figure}[t]
     \includegraphics[scale=0.45]{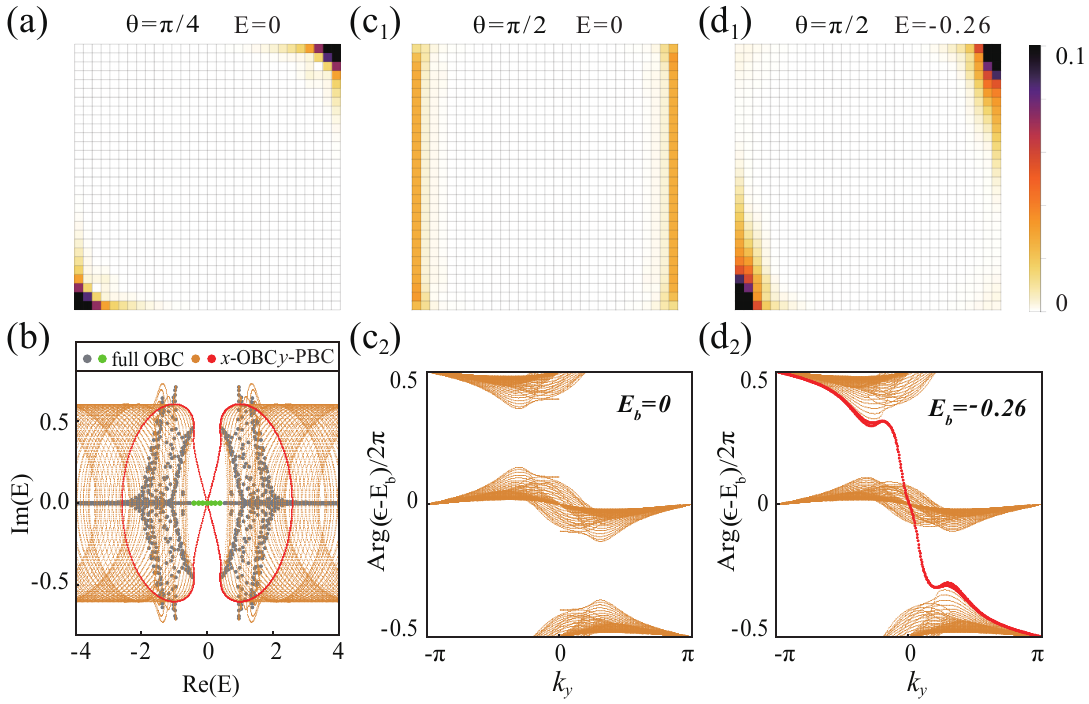}
     \caption{The correspondence between winding number $(\mathcal{W}_x,\mathcal{W}_y)$ and the skin-topological modes with Zeeman field. (a)Eigenstate distribution of the zero modes with $\theta=\frac{\pi}{4}$. (b)The energy spectrum for non-Hermitian Hamiltonian $H_1(\mathbf{k})$ under the full OBC for the bulk (gray) and surface (green) states, the only OBC in the $x$ direction (orange and orange dots), with $\theta=\pi/2$. The green dots are edge modes. ($\mathrm{c}_1$)-($\mathrm{c}_2$) Eigenstate distribution and phase plots of $E=0$. ($\mathrm{d}_1$)-($\mathrm{d}_2$) Eigenstate distribution and phase plots of $E=-0.26$. Values of parameters are $\mu=3, t=1, \Delta=1, V=0.5, \delta_x=0.3, \delta_y=0.3$}\label{fig4}
\end{figure}

Upon the introduction of a Zeeman field, the position of MZM will change with the rotation of the Zeeman field. When $\theta=\pi/2$ or $\pi$, the edge states with zero energy will be uniformly distributed on the edge, while the edge states with non-zero energy are subject to the influence of the $Z_2$ NHSE and localized on the corner, as shown in Fig. \ref{fig4}($\mathrm{c}_1$)-($\mathrm{d}_1$). The energy under $x$-OBC and $y$-PBC (red dots) encircle the energy of the skin-topological modes with non-zero energy under all OBC(green dots) in Fig. \ref{fig4}(b). The winding number of the edge mode with zero energy is zero, so it is not affected by the $Z_2$ NHSE, as illustrated in Fig. \ref{fig4}($\mathrm{c}_2$). The winding number $\mathcal{W}_x$ of the topological edge state with non-zero energy is depicted in Fig. \ref{fig4}($\mathrm{d}_2$).

\subsection{\textnormal{\textbf{spin-dependent nonreciprocal hopping}}}\label{Sec:IIIB}

\begin{figure}[t]
     \includegraphics[scale=0.65]{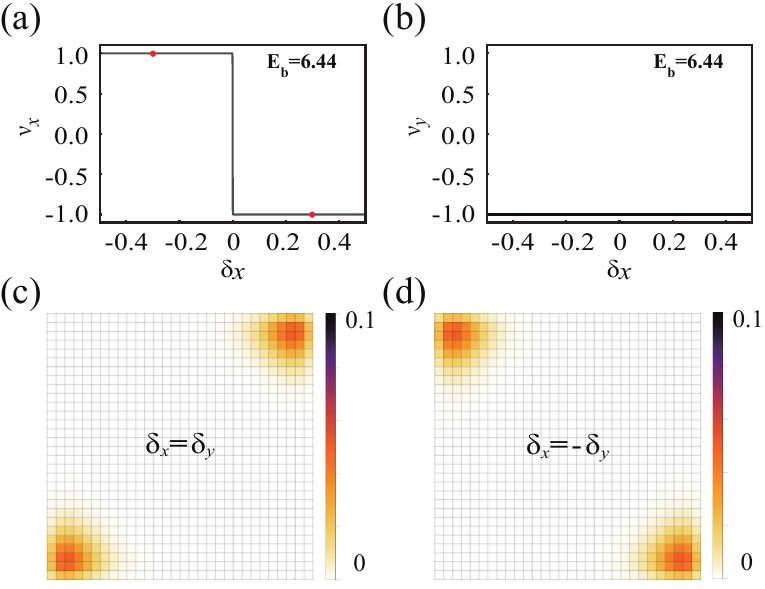}
     \caption{The correspondence between winding number $(\nu_x,\nu_y)$ and the position of the skin modes in bulk without Zeeman field. (a)-(b)Topological invariant $\nu_\alpha$ with varied $\delta_x$ under the only PBC in $\alpha$ direction, red dots are $\delta_x=\delta_y=0.3$ or $\delta_x=-\delta_y=-0.3$, in which $k_y=0$. (c)-(d) Eigenstate distribution of the bulk modes with $\delta_x=\delta_y=0.3$ in (c) and $\delta_x=-\delta_y=-0.3$ in (d).The rest parameters are determined as $\mu=3, t=1, \Delta=1,\delta_y=0.3, V=0$. }\label{fig5}
\end{figure}

With the addition of spin-dependent nonreciprocal hopping as in Eq. (\ref{ham:eq2}), in the absence of Zeeman field, the bulk eigenstates and topological edge modes exhibit non-Hermitian skin effect and are located in the corners of the system. In this case, the system possesses TRS, which dictates that each complex eigenvalue comprises a Kramers pair positioned diagonally. Next, we will analyze the characteristics of $Z_2$ NHSE protected by TRS.

The eigen-equation of the tight-binding Hamiltonian H of the system is
\begin{equation}
	H|\psi\rangle=E|\psi\rangle, \quad  H^\dagger|\chi\rangle=E^*|\chi\rangle
\end{equation}
in which $|\psi\rangle$ and $|\chi\rangle$ are the right and left eigenstates, respectively. The left eigenstate $|\chi\rangle$ satisfies 
\begin{equation}
	H(T|\chi\rangle^*)=TH^T|\chi\rangle^*=E(T|\chi\rangle^*)
\end{equation}
which means that both $|\psi\rangle$ and $T|\chi\rangle^*$ are right eigenstates of the energy E, and are orthogonal and linearly independent of each other \cite{Kawabata2020May}. Due to the periodicity of space, the eigenstates can be written as linear combinations $|\psi\rangle=\sum_{j}\beta_{j}^{n}\phi_{j}$. The characteristic equation of the non-Bloch Hamiltonian can be written as
\begin{equation}
	H(\beta_j)|\phi_j\rangle=E|\phi_j\rangle.
\end{equation} Because of TRS, we have
\begin{equation}
	H(\beta_{j}^{-1})T|\chi_j\rangle^*=TH(\beta_j)^{T}|\chi_j\rangle^*=ET|\chi_j\rangle^*
\end{equation}
This indicates that $|\phi_j\rangle$ and $T|\chi_j\rangle^*$ form a Kramer pair, which are localized on opposite sides of the system.

\begin{figure}[t]
     \includegraphics[scale=0.45]{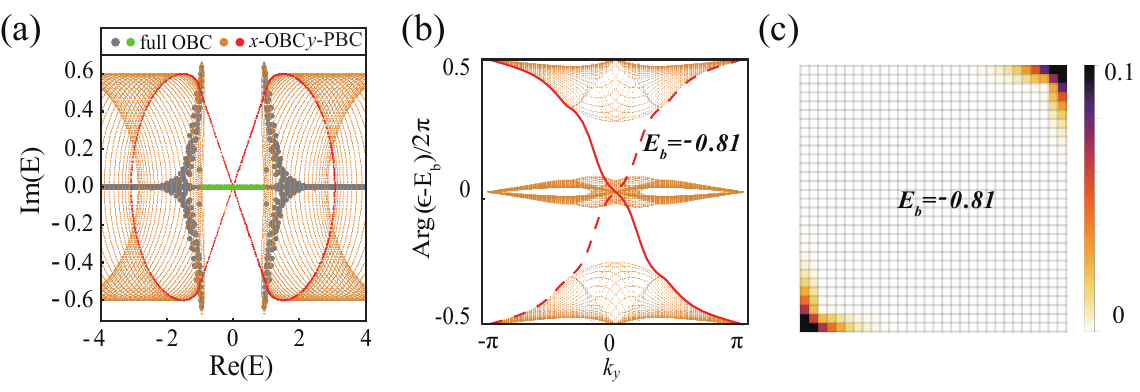}
     \caption{The correspondence between winding number $(\mathcal{W}_x,\mathcal{W}_y)$ and the position of the skin-topological modes without Zeeman field. (a)The energy spectrum for non-Hermitian Hamiltonian $H_2(\mathbf{k})$ under the full OBC(gray dots) and the only OBC in the $x$ direction (orange dots), with $V=0$. The green dots are edge modes.(b) Phase plots of the eigenvalues with respect to the wave vector $k_y$. The reference value is $E_b=-0.81$. (c)Eigenstate distribution of the edge modes.}\label{fig6}
\end{figure}

The winding number in Eq. (\ref{W:eq3}) can be rewritten as
\begin{align}
	W_\alpha(k_\perp,E_b)&=\sum_{n=1}^{2N}\int^{2\pi}_{0}\frac{dk_\alpha}{2\pi i}\frac{\partial}{\partial k_\alpha}\log[E_n(k_\alpha,k_\perp)-E_b]\nonumber\\
	&=\sum_{n=1}^{2N}W_\alpha^n(k_\perp,E_b)
\end{align}
Where $E_n(k_\alpha,k_\perp)$ is the $n$th band of $H_{2}(\mathbf{k})$ and $W_\alpha^n(k_\perp,E_b)$ is the winding number of the $n$th bands. If the $p$th and $q$th energy bands are TRS of each other, then their winding number satisfies the formula $W_\alpha^p(k_\perp,E_b)=-W_\alpha^q(k_\perp,E_b)$. Then the topological invariant of a non-Hermitian system with TRS is 
\begin{equation}
	\nu_\alpha                                                                                                                                                                                                                                                                                                                                                                                                                                                                                                                                                                                                                                                                                                                                                                                                                                                                                                                                                                                                                                                                                                                                                                                            (k_\perp,E_b)=\frac{1}{2}\sum_{n=0}^{N-1}|W_\alpha^{2n+1}(k_\perp,E_b)-W_\alpha^{2n+2}(k_\perp,E_b)|\label{v:eq10}
\end{equation}
where $(2n+1)$th and $(2n+2)$th bands are TRS of each other \cite{Wang2024}. We show the topological invariant $\nu_\alpha$ change with $\delta_x$ in Fig. \ref{fig5}(a)-(b). If $\nu_\alpha\neq 0$, then the bulk eigenstate of $E_b$ is affected by $Z_2$ NHSE protected by TRS. If the sign of $\nu_x\nu_y$ is positive, bulk modes are localized in the lower left and upper right corners; while if the sign of $\nu_x\nu_y$ is negative, they are in the upper left and lower right corners, as depicted in Fig. \ref{fig5}(c)-(d).

The edge states are also affected by $Z_2$ NHSE. The energy under $x$-OBC and $y$-PBC (red dots) encompasses the energy of the edge states under full OBC(green dots). We calculate the winding number $\mathcal{W}_\alpha$ of the topological edge states in Fig. \ref{fig6}(b). Because of $Z_2$ NHSE, the edge states are localized in the corner, as show in Fig. \ref{fig6}(c). TRS results in the existence of two eigenstates with identical energy E, forming a Kramers pair and these eigenstates exhibit opposite winding numbers.
\begin{figure}[t]
	\includegraphics[scale=0.45]{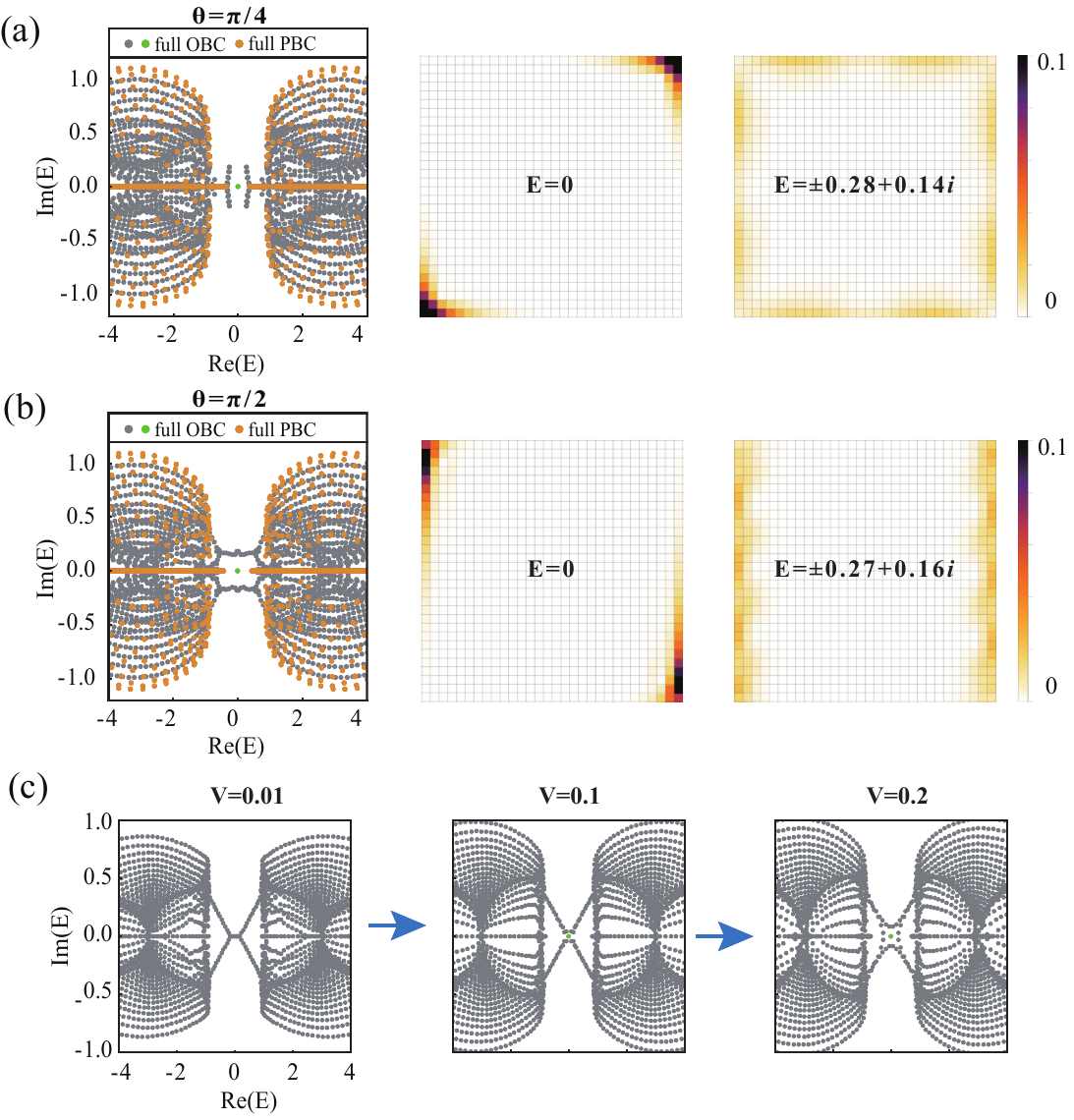}
	\caption{(a)-(b)The energy spectrum, distribution of MZM and edge states when $\theta=\frac{\pi}{4}$(a) and $\theta=\frac{\pi}{2}$(b). (c)The energy spectrum varies with the Zeeman field, in which $\theta=\frac{\pi}{2}$. The rest parameters are determined as $\mu=3, t=1, \Delta=1,\delta_x=0.3, \delta_y=0.3, V=0.5$.}\label{fig7}
\end{figure}

 When the Zeeman field is added, the system changes from symplectic class to orthogonal class, where the eigenvalue no longer has Kramers degeneracy. For the orthogonal class, the characteristic equation reads
 \begin{align}
	\det[H_2(\bm{\beta}^{-1})-E]&=\det[T'H^{T}_{2}(\bm{\beta})T'^{-1}-E]\nonumber  \\
	&=\det[H_2(\bm{\beta})-E]\nonumber\\
	&=0,
\end{align}
meaning no NHSE can exists for the continuous energy bands. However, for the isolated energy states, it  might still be localized \cite{Kawabata2020May}.  Indeed as shown in the middle columns of Fig. \ref{fig7}(a) and Fig. \ref{fig7}(b), the isolated zero energy states are still affected by the NHSE and localized in the corner. In right columns of Fig. \ref{fig7}(a) and Fig. \ref{fig7}(b), the edge states in continuous energy band have no NHSE and are uniformly distributed along the edge. When increasing the external Zeeman field as shown in Fig. \ref{fig7}(c), the energy spectrum under OBC undergo topological phase transition, and MZM appears at $V=0.1$. When taking the thermodynamic limit, given arbitrary small Zeeman fields, the energy spectrum under OBC will undergo significant variations.\cite{Okuma2019}.

\section{SUMMARY}\label{Sec.IV}
In conclusion, we reveal the topological properties and non-Hermitian skin effects of two-dimensional topological superconductor in two different nonreciprocal systems. For spin-independent nonreciprocal hopping, the bulk eigenstates exhibit $Z_2$ NHSE because of PHS. We utilize the winding number as an indicator to pinpoint the localization direction of NHSE. At the same time, $Z_2$ NHSE also acts on the edge states so that the edge states are localized at corners. When the Zeeman field is added, the position of MZMs will also shift with the change of the angle of the Zeeman field. At $\theta=\frac{\pi}{2}$, the edge states with zero energy are uniformly distributed on the opposite sides, while the edge states with non-zero energy is affected by $Z_2$ NHSE and localized on the opposite corners. For spin-dependent nonreciprocal transitions, $Z_2$ NHSE is protected by TRS, and the eigenstates of bulk are also localized in two opposite corners. We use topological invariant $\nu_\alpha$ to characterize this $Z_2$ NHSE, and the edge states are also affected by $Z_2$ NHSE, thus localized in the opposite corner. Due to the nature of TRS protection, NHSE is suppressed upon the introduction of the Zeeman field for the states in the continuum, while survived for the isolated zero energy states, and topological phase transition occurs as the Zeeman field increases. Our work has further deepened the understanding of symmetry-protected NHSE and provided a new platform for the study of the interplay between topological properties and non-Hermiticity.

\begin{acknowledgments}
This work is financially supported by National Natural Science Foundation of China  No. 92265103, the National Basic Research Program of China (Grants No. 2015CB921102 and No. 2019YFA0308403), and the Innovation Program for Quantum Science and Technology (Grant No. 2021ZD0302400).
\end{acknowledgments}

\bibliography{NHSOTSC.bib}
\onecolumngrid
\end{CJK}
\end{document}